\documentclass{article}
\usepackage{amssymb}
\usepackage{graphicx}
\usepackage[dvips]{color}
\title{Monopole clusters, center vortices, and confinement in a
Z(2) gauge-Higgs system }
\author{F.Gliozzi and A. Rago}
\date{June 2002}
\newcommand{\eq}{\begin{equation}}
\newcommand{\en}{\end{equation}}
\newcommand{\ear}{\begin{eqnarray}}
\newcommand{\rae}{\end{eqnarray}}
\newcommand{\Z}{\mathbb{Z}}
\newcommand{\bra}{\langle}
\newcommand{\ket}{\rangle}
\newcommand{\um}{\frac12}
\definecolor{M_Beige}         {rgb}{0.96 , 0.96 , 0.86}

\definecolor{M_Brown}         {rgb}{0.65 , 0.16 , 0.16}

\definecolor{M_Gold}          {rgb}{1.00 , 0.84 , 0.00}

\definecolor{M_LemonChiffon}  {rgb}{1.00 , 0.98 , 0.80}

\definecolor{M_Orange}        {rgb}{1.00 , 0.60 , 0.00}

\definecolor{M_Pink}          {rgb}{1.00 , 0.75 , 0.80}

\definecolor{M_Violet}        {rgb}{0.93 , 0.51 , 0.93}
\begin{document}
\maketitle
\noindent
 Dipartimento di Fisica Teorica, Universit\`a di Torino and\\ INFN,
sezione di Torino, via P. Giuria, 1, I-10125 Torino, Italy.
\vskip0.5cm\noindent
e-mail: gliozzi, rago@to.infn.it
\begin{abstract}
We propose to use the different kinds of vacua of the gauge theories 
coupled to matter as a laboratory to test  confinement ideas 
of pure Yang-Mills theories. In particular, the very poor overlap of
the Wilson loop with the broken string states supports the 't Hooft and
Mandelstam confinement criteria. However in the $\Z_2$ gauge-Higgs model
we use as a guide we find that the condensation of monopoles and 
center vortices is a necessary, but not sufficient condition for confinement.
\end{abstract}
\section{Introduction}
Center vortices \cite{th} and magnetic monopoles \cite{tm} are widely
believed to be the collective degrees of freedom responsible for the
non-perturbative features of $SU(N)$ Yang Mills theories, in
particular for confinement. 

Condensation of magnetic monopoles implies a dual Meissner effect:
the chromolelectric field is expelled from the vacuum and gives 
the well-known physical picture of confinement 
in terms of dual Abrikosov vortices which describe the flux tubes
joining static sources.

Center vortices are string-like excitations formed out of  the center
of the gauge group. They produce a very efficient disordering
mechanism of the gauge configurations which could  lead to area law decay
of large Wilson loops.

Although the above descriptions for almost all known models
remains at a conjectural stage\footnote{The only explicit, analytic, example 
of these mechanisms can be found in the Polyakov \cite{po} proof of confinement
of compact $U(1)$ gauge model in 2+1 dimensions.}, in the last few
years many numerical lattice studies have
given strong support for the relevant role played by center vortices 
and monopole condensation in confinement.  
For instance in the $SU(2)$ gauge model in 3+1 dimensions it has 
been observed both  magnetic monopole condensation 
\cite{Nakamura:1996sw,Chernodub:1996ps,ht,DiGiacomo:1999fa}
and the phenomenon of center dominance \cite{DelDebbio:1996mh}, 
namely the fact that the string tension obtained from center projected
configurations in maximal center gauge 
agrees with the same quantity calculated in the full theory.

Center dominance is verified in a trivial way in abelian theories, where the
whole dynamics is described by center vortices. Here the distinction
between confined and unconfined phase is related to the maximal size of
the clusters of vortices: it can be shown that the confinement
requires the presence of an infinite cluster
\cite{Gliozzi:2002ht}. This agrees with an earlier observation of 
 percolating  center vortices in the cold phase of
$SU(2)$ gauge model at finite temperature \cite{Engelhardt:1999fd}.
For the sake of brevity we call vortex condensation the appearance of such a 
kind of infinite cluster.
 
There are many open, intertwined, questions about the actual validity
of the 't Hooft criteria of confinement, namely that the 
monopole and/or vortex condensations imply the decay with an area law
of large Wilson loops. In particular, is it possible to derive monopole 
condensation from vortex condensate or vice versa ? If not, are both
condensations necessary for confinement? Are they also sufficient?

In this work we propose to answer these questions 
by studying gauge systems coupled to matter.  
Indeed it is worth noting that while center vortices and monopoles 
are generally studied in pure Yang Mills models, they are well defined 
also in gauge theories coupled to matter. In these models there are 
different vacua, distinguished by different entities which condense.
If, for instance, the region where the vortices condense does not
coincide with that with monopole condensation one can infer that these 
two properties are logically independent. This is precisely what happens
in a 2+1 $\Z_2$ gauge-Higgs model \cite{Gliozzi:2001tu} .  
A simple argument shows \cite{Gliozzi:2002am} that the only  vacua 
compatible with an area-law decay of the large Wilson loops
 are those in which both magnetic monopoles \underline{and}  center
 vortices condense. In other
terms, the 't Hooft criteria are both necessary for confinement. 

Assuming that they are also sufficient can explain a surprising
phenomenon observed in almost all  coupled gauge systems studied up to
now: although 
the potential between static sources flattens at large distances because
of the screening produced by pair creation, this flattening (called
string breaking) is invisible in the Wilson loop: it continues to obey
an area law   in full QCD \cite{qcd}  even at
distances where the static charges are completely screened.
The point is that in gauge theories coupled to matter the basis of the 
operators has to be enlarged \cite{Michael:1991nc} in order to get a reliable
estimate of the potential. In this way it has been observed the
breaking of the confining string in Higgs models \cite{ks} and in QCD 
\cite{milc}. 
So the fact that large Wilson loops obey an area law even  in
coupled systems, as first suggested in \cite{gp}, may be considered
as a further support to the usual  plausibility arguments for the 
confinement mechanisms. Of course 
numerical experiments cannot give a true proof of the sufficiency of
these criteria: it could well happen that Wilson loops of much larger
size exhibit  string breaking explicitly.   

In this work we study the 2+1 $\Z_2$ gauge-Higgs model to investigate
a subtler issue:by probing the different
vacua characterised by infinite vortex and monopole clusters  we study 
the effect of other  condensates on confinement.
In this model there are two  vacua having
center vortex and magnetic monopole condensates; one of them has
also an electric condensate, i.e. the Higgs field has a vacuum
expectation value different from zero.

The former fulfils all the requirements of the confinement
criteria, so an area law is expected; here we  find that  the
Wilson loop obeys a perfect area law even at distances larger than
five times the string breaking scale.

The latter  can be identified with the torn phase predicted in 
Ref.\cite{gp}: the Wilson loop decays with an area law below a given
scale which varies very rapidly as a function of the couplings of the
model and is unrelated to the string breaking scale. Above this
threshold we observe a perimeter law: the infrared properties of this vacuum
are indistinguishable from that of the perturbative unconfined
vacuum\footnote{A similar phase  has been reported  in 4D
  $SU(2)$-Higgs model \cite{Bertle:2001ya}}.

\section{The model}
The action of a  3D $\Z_2$ gauge theory coupled to  
 matter in a cubic lattice $\Lambda$ can be written as 
\eq
S(\beta_G,\beta_I)=-\beta_I\sum_{\langle ij\rangle}
\varphi_i U_{ij}\varphi_j-
\beta_G\sum_{plaq.}U_{\square}~,
\en
where both the link variable $U_{ij}\equiv U_\ell$  and the matter
field $\varphi_i$  take values $\pm1$ and
$U_{\square}=\prod_{\ell\in\square}U_\ell$.

This model is self-dual: the Kramers- Wannier transformation maps the
model into itself. Its partition function
\eq
Z(\beta_G,\beta_I)=\sum_{\{\varphi_i=\pm1,~U_\ell=\pm1\}}
e^{-S(\beta_G,\beta_I)}\en
fulfils  the functional equation
\eq
Z(\beta_G,\beta_I)=(\sinh 2\beta_G \sinh 2\beta_I)^{\frac 32 N}
Z(\tilde{\beta}_I,\tilde{\beta}_G)
\en
with   $\tilde\beta=-\um\log(\tanh \beta)$.

\begin{figure}
%\begin{center}
%\input{largest.tex}
%\end{center}
\centering
\includegraphics[width=0.8\textwidth]{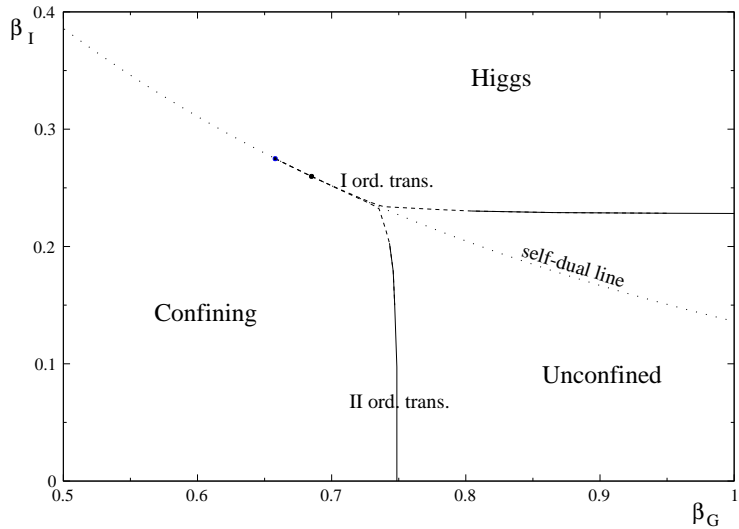} 
%\vspace{5cm}  % amount of vertical space needed
\caption{Phase diagram for the 2+1 $\Z_2$ gauge-Higgs system. The
  dashed (solid) lines denote the first-order (second-order)
  transitions. The dotted line is the self-dual line. }
\label{fig:phase_diag}
\end{figure}

The phase diagram of this model (see Fig. \ref{fig:phase_diag}) has
been studied long ago
\cite{js}. There is an unconfined region surrounded by lines of phase
transitions toward  the Higgs phase and its dual.
These lines are second order until they are near each other and the
self-dual line, where first order transition occurs  \footnote{For a
more detailed description of the phase diagram see the poster
presented by A.Rago at Lattice 2002.}. 

\subsection{ $\Z_2$ vortices and monopoles}
In this model the construction of center vortex configurations is 
straightforward: to each frustrated plaquette
(i.e. $U_{\square}=-1$) assign a vortex in the dual link.  Since
the product of the plaquettes belonging to any elementary cube is 1,
center vortices form closed subgraphs of even coordination number.
Thus a connected vortex subgraph contributes to a given Wilson loop
 $W(C)$ only if an {\sl odd} number of lines are linked to it: 
the Wilson loop is a vortex counter modulo 2. 

Magnetic monopoles can be defined exactly as in pure gauge
model. They live on the dual lattice $\tilde{\Lambda}$. To create a 
monopole in a site $\tilde{x}\in\tilde{\Lambda}$,
corresponding to the center of an elementary cube of $\Lambda$, it is
sufficient to draw an arbitrary, continuous line $\gamma(\tilde
x,\tilde y)$ 
joining $\tilde{x}$ to  to another monopole located in $\tilde y$ (or
to $\infty$) and flipping the sign of the coupling  of the plaquettes 
crossed by $\gamma$. This flipping is generated by the
non-local operator
\begin{equation} 
\Psi_\gamma(\tilde x,\tilde y)=\exp(-2\beta_G\sum_{\square\in\gamma}U_\square)~~.
\end{equation} 
As a consequence, the flux across any closed
surface with $\tilde{x}$ inside and $\tilde y$ outside is equal to -1: 
this monopole field
$\Psi_\gamma(\tilde x,\tilde y)$ creates one unity of  $\Z_2$ flux
joining $\tilde x$ to $\tilde y$.
Monopole condensation occurs when
\eq
\lim_{|\tilde x-\tilde y|\to\infty}\bra\Psi_\gamma
(\tilde x,\tilde y)\ket\not=0~~~.
\en  
Why this condensation should imply confinement?
A useful piece of information comes from the Kramers- Wannier duality.
 One can easily show that under this duality map the monopole 
correlator transforms as follows 
\eq
\bra\Psi_\gamma(\tilde x,\tilde y)\ket_{\beta_G,\beta_I}=
\bra\varphi_{\tilde x}\prod_{\ell\in\gamma}U_\ell\,\varphi_{\tilde y}
\,\ket_{\tilde\beta_I,\tilde\beta_G}~~.
\en
Thus the monopole condensation is associated to the dual Higgs phase,
where  the $\Z_2$  symmetry is spontaneously broken.

The same transformation maps the Wilson loop $W(C)$ associated
 to any closed curve $C\in\Lambda$ into the corresponding 't Hooft loop 
$\widetilde{W}(C)$ of the dual phase:
\eq
\bra W(C)\ket_{\beta_G,\beta_I}= \bra \widetilde{W}(C)
\ket_{\widetilde{\beta}_I,\widetilde{\beta}_G}
\label{dual}
\en
with
\eq
\widetilde{W}(C)=\exp(-2\widetilde{\beta}_G\sum_{\bra ij\ket\in\Sigma}
\varphi_i\,U_{ij}\,\varphi_j)~~,
~\; \partial\Sigma=C
\en
where $\Sigma$ is an arbitrary surface bounded by $C$.
$\widetilde{W}(C)$ creates an elementary $\Z_2$ flux along $C$ which 
manifests itself as a topological defect: the action of
$\widetilde{W}(C)$ on an arbitrary configuration maps
the product $\eta_{\widetilde C}=\prod_{\ell\in\widetilde C} U_\ell$ 
along any loop $\widetilde{C}\in\widetilde{\Lambda}$ having 
an odd linking number with $C$ into $-\eta_{\widetilde C}$.

\subsection{A microscopic picture of confinement}
Confinement at $\beta_G,\beta_I$ requires  area law decay of 
$\bra\widetilde{W}(C)\ket_{\widetilde{\beta_I}\widetilde{\beta_G}}$.
To understand this property at a microscopic level it is convenient to 
resort to the  Fortuin Kasteleyn (FK) random cluster representation
\cite{fk} of the model. Starting from the obvious identity
\eq
{\rm e}^{\beta_I\varphi_xU_{xy}\varphi_{y}}={\rm e}^{\beta_I}(1-p+p\,
\delta_{1,\varphi_xU_{xy}\varphi_y})~,~ p=1-{\rm e}^{-2\beta_I}
\en
It is easy to perform explicitly the sum on the matter fields
$\varphi_x$, which yields
 
\eq
Z(\beta_G,\beta_I)=\sum_{U_\ell}{\rm e}^{\sum_\square\beta_G U_\square}
 \sum_{G\subseteq\Lambda}\varpi_U(G)\,v^{b_G}2^{c_G}~, 
~~v=\frac{p}{1-p},
\label{Z}
\en
The summation is over all 
spanning subgraphs $G\subseteq\Lambda$. $b_G$ is the number of links of $G$, 
called active bonds (which are the matter degrees of freedom replacing
the $\varphi$'s in this representation),
and $c_G$ is the number of connected components, called FK clusters. 
%The ordered phase
% is characterised by the presence of an infinite, percolating  FK 
%cluster.
$\varpi_U(G)$ is a projector on the subgraphs which are
compatible with a given gauge configuration $U=\{U_\ell\}$:
%When there are frustrations in the system (for instance those generated 
%by $\widetilde{W}(C)$)
% the summation in Eq.\ref{Z} is constrained. 
only those subgraphs are 
allowed for which no closed path within each FK cluster is linked to an 
elementary $\Z_2$ flux \cite{Caselle:1999hy}. Put differently, no
frustration is permitted in $G$. In a sense, this is a 
microscopic realization of a sort of dual Meissner effect: the FK 
clusters behave like
pieces of dual superconducting matter of type I and no $\Z_2$ flux can 
go through the circuits of $G$. This is the only constraint generated
by the interaction
between matter  and gauge fields. In the limit $\beta_G\to\infty$ all
the plaquettes have $U_\square=1$ (trivial gauge vacuum) hence the sum over
the subgraphs is unconstrained and one gets the standard Ising model. 
 \begin{figure}[t]
%\begin{center}
%\input{largest.tex}
%\end{center}
\centering
\includegraphics[width=0.8\textwidth]{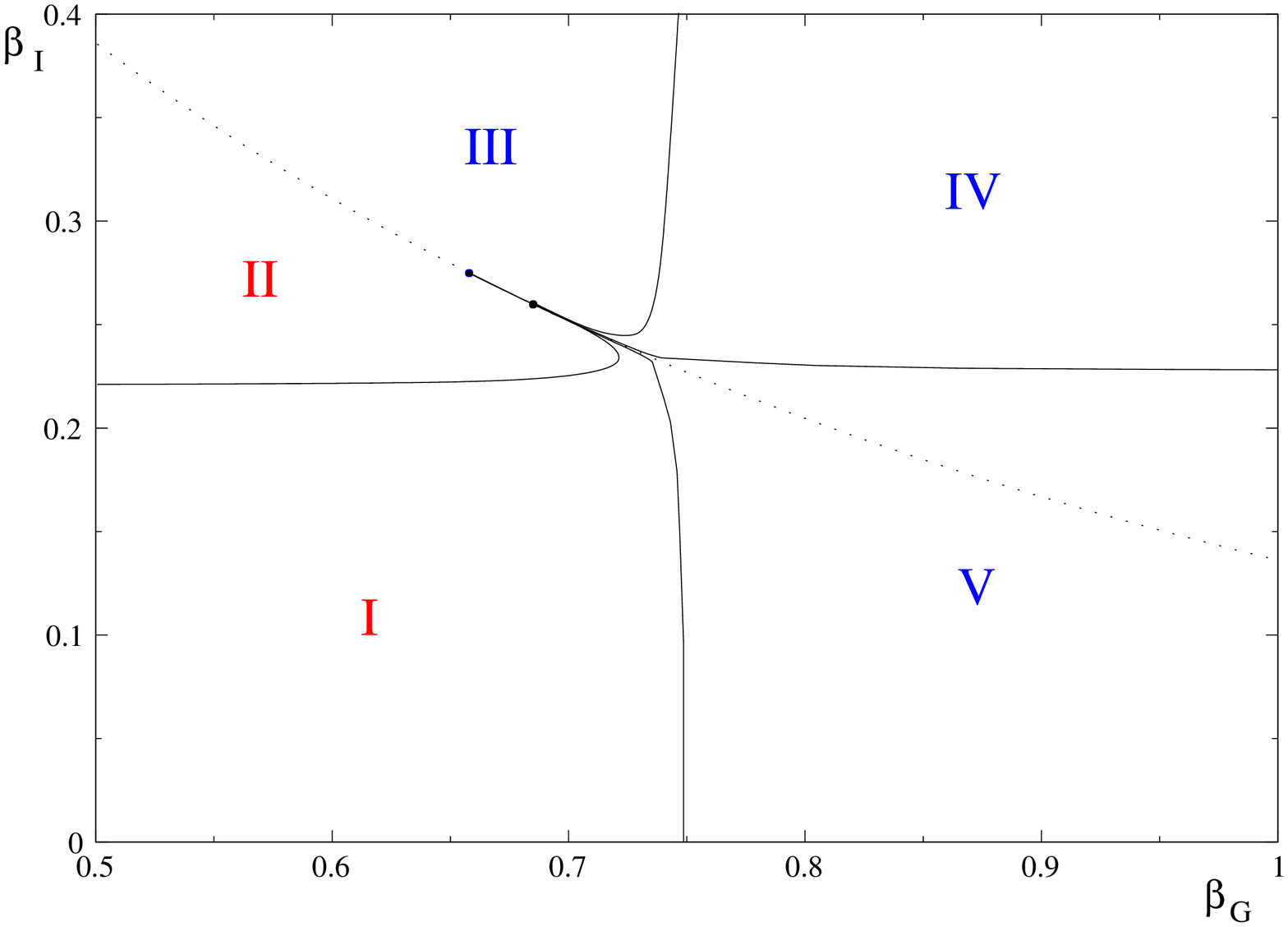} 
%\vspace{5cm}  % amount of vertical space needed
\caption{The different vacua of the $\Z_2$ gauge-Higgs model,
  characterised by different kinds of infinite FK or CV clusters 
(see Tab.\ref{tab:vacua}).}
\label{fig:vacua}
\end{figure}
%\vskip -1. cm 
Introducing a further projector $\varpi_C(G)$ on the space of configurations 
$\{G\subseteq\Lambda\}$, which takes 
the value 0 whenever there is  a FK cluster (at least)  linked to 
$C$, and the value 1  in all the other cases, yields the very useful identity
\eq
\bra\widetilde{W}(C)\ket=\bra\varpi_C\ket=\frac{\rm number~ of~ compatible~ 
config.}{\rm total~number~ of~ config.}~~.
\label{wc}
\en 
Assume that there are only FK clusters of finite size ( this is the 
case of the region V in Fig. \ref{fig:vacua}). If $C$ is much 
larger than the mean 
size of the clusters, 
the configurations contributing to $W(C)$, i.e. those with $\varpi_C(G)
=1$, are
 characterised by the fact that there is no cluster linked to $C$. The 
weight of this class
of configurations, when compared with the total ensemble 
$G\subseteq\Lambda$ is 
clearly suppressed by a factor $e^{-\alpha\, p(C)}$, where $p(C)$ is the 
length of $C$. Thus in this phase the Wilson loop obeys a perimeter law. 

Conversely, a large Wilson loop decaying with an  area law
implies by necessity the presence of an infinite  FK cluster. 
In order to see whether this condition is also sufficient for confinement 
in the dual phase, note that according to Eq.(\ref{wc}) an infinite FK 
cluster $fk_\infty$ gives a non-vanishing contribution to  $\widetilde{W}(C)$ 
only if there is at least a simply connected surface $\Sigma$ bounded
by $C$ which is
not pierced by the loops of the cluster.
If the links of $fk_\infty$ are weakly correlated, one is led to argue
that the weight of the configurations compatible with
$\widetilde{W}(C)$ 
is suppressed by a factor ${\rm e}^{-\sigma\,a(\Sigma)}$, where
$a(\Sigma)$ is the area of the minimal surface with
$\partial\Sigma=C$. 
This leads to the area-law decay of $\bra\widetilde{W}(C)\ket$.
\par
A crucial assumption in the above argument is the weak correlation of
the links of $fk_\infty$ which describe at the microscopic level the
monopole condensation. We shall see that there is a phase in which,
though such a condensation occurs, this assumption is no longer
true. Correspondingly we observe a violation of the area-law.
 
\par
We can now apply the same line of reasoning to the center
vortices (CV). Finite CV clusters  can link with the 't Hooft loop  only
along its perimeter. Therefore they contribute only to the perimeter
term.  
An infinite CV cluster is necessary for the area-law:
confinement implies both infinite center
vortex subgraph \underline{and} magnetic monopole condensation.
In pure gauge theory these two requirements coincide, while in the 
coupled system the vacuum structure is more intriguing.    

\subsection{The vacua} 

In the coupled theory we have two kinds  of dynamical subgraphs:
CV  {\sl or}  FK clusters in the \underline{dual} 
lattice describe the gauge field degrees of freedom; FK clusters in the
\underline{direct} lattice describe the charged Higgs
matter. 

In order to recognise an infinite cluster in practical simulations 
one has to look at the cluster size  $s$ as a function of the
lattice volume $V$. We say that there is an infinite cluster whenever
$s\propto V$ for large enough $V$. 
Straightforward numerical experiments show that 
 they are distributed in the phase diagram  according to
Fig.\ref{fig:vacua} and Tab.\ref{tab:vacua}.

There are four kinds of infinite clusters: {\sl i)} FK cluster in the
dual lattice (magnetic condensate), {\sl ii)} FK cluster in the direct
lattice (electric condensate), {\sl iii)} CV cluster  in the
dual lattice, {\sl iv)} CV cluster in the direct lattice (dual CV).

In the region V there is no infinite cluster of any type. This is
the perturbative, weak coupling vacuum, where  large Wilson
loops \underline{and} 't Hooft loops obey a perimeter-law.

The region IV is characterised by an electric condensate: it is the
normal Higgs phase, where the Wilson loops decay with a perimeter-law,
while the 't Hooft loops obey an area-law.
 
\begin{table}
\caption{Distribution of the infinite clusters in the phase diagram}
\vskip .2 cm
\centering
\begin{tabular}{ccccc}
\hline
phase& magnetic& electric& center&dual\\
~& condensate&condensate& vortices& center vortices\\
\hline
I&yes&no&yes&no\\
II&yes&yes&yes&no\\
III&yes&yes&no&yes\\
IV&no&yes&no&yes\\
V&no&no&no&no\\
\hline
\end{tabular}
\label{tab:vacua}
\end{table}
%\vskip .5 cm

Region III has a magnetic condensate but no large center vortices, then 
there is no confinement, as confirmed by numerical tests. This region
is dual to the region II, where we shall see that the Wilson loops
follow a perimeter-law decay. Because of duality relation (\ref{dual}) we
can infer that also $\widetilde{W}(C)$ decays in the same way.

The regions
I and II have both CV and FK infinite clusters in the dual lattice. 

The former is a normal confining phase: the potential extracted from
an enlarged basis shows the expected string breaking 
(see Fig.\ref{fig:enl_basis}b), 
while the Wilson loop obeys a perfect area-law even at a large scale
(see Fig.\ref{fig:wilson_loop_fk}). This has been  checked up to
distances of the order of five times the string breaking scale (compare
Fig.\ref{fig:enl_basis}b and Fig.\ref{fig:wilson_loop_fk}). 
 
The latter is a phase with the simultaneous presence of
infinite  clusters in the direct and the dual lattices. 
The infinite cluster of $\Lambda$ is associated to the
condensation of the Higgs field $\varphi_x$, while the infinite CV
cluster of $\widetilde{\Lambda}$ is typical of a confining
phase. However there is no confinement in the IR limit:  the
no-frustration constraint induces strong correlations among $\Z_2$
flux lines, because only an even number of them can pass through
the closet paths within the FK clusters. 

As a matter of fact, large
Wilson loops obey a perimeter-law decay, even if at intermediate
distances an area-law seems recovered. A first example of this behavior
is reported in \cite{Gliozzi:2001tu}. The cross-over scale varies rapidly
as a function of the coupling constant and is unrelated to the string 
breaking scale.

\section{Numerical results}

\begin{figure}
\centering
\begin{minipage}[t]{.48\textwidth}
\centering
\includegraphics[width=0.99\textwidth]{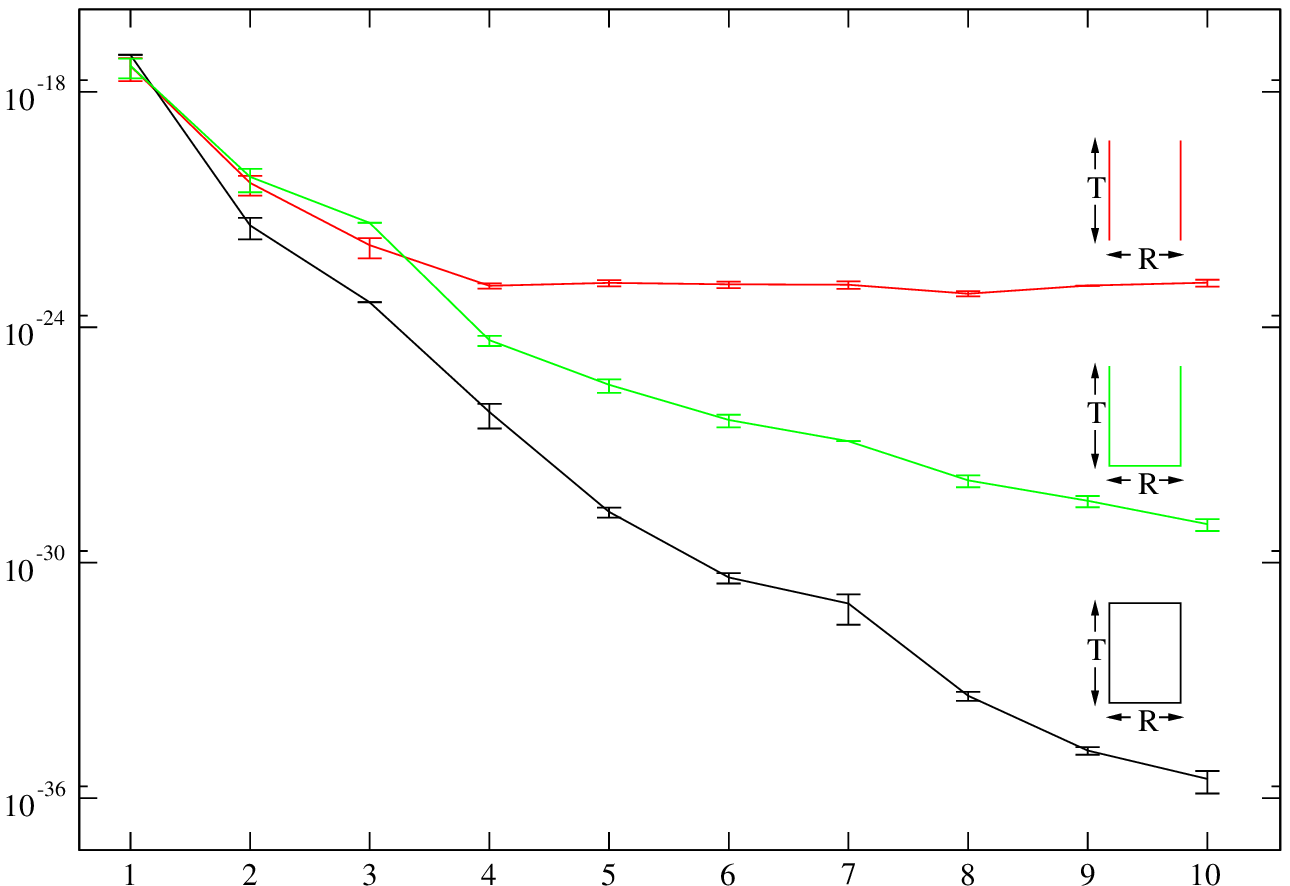}\\
\centerline{ (a)}
\end{minipage}
\hspace{.02\textwidth}
\begin{minipage}[t]{.48\textwidth}
\centering
\includegraphics[width=0.98\textwidth]{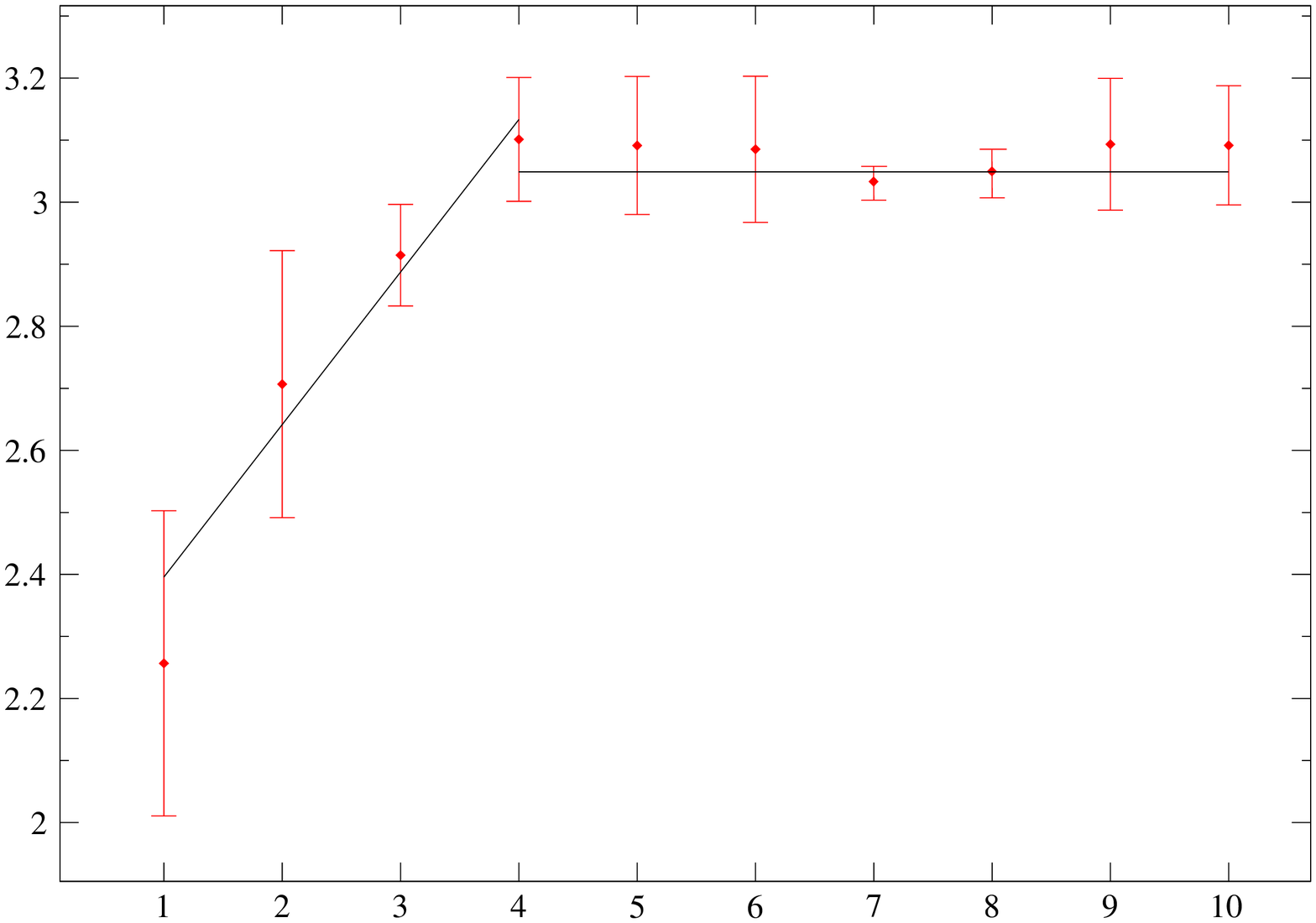}\\
\centerline{ (b)}
\end{minipage}
\caption{Results of Monte Carlo simulations of $6.0 \cdot 10^{5}$
  sweep of update on cubic lattice of size $40^3$:
(a):Plot of the rectangular Wilson $W(R,T)$ loop and the other two
  operators defined in Eq.s (\ref{op1}) and (\ref{op2}) at $T=17$ 
 as a function of $R$.
(b):The static potential extracted by the lowest eigenvalue of the
correlation matrix obtained for 6 different values of $T$
($T=15,\cdots,20$)}
\label{fig:enl_basis}
\end{figure}
\begin{figure}
%\begin{center}
%\input{largest.tex}
%\end{center}
\centering
\includegraphics[width=.9\textwidth]{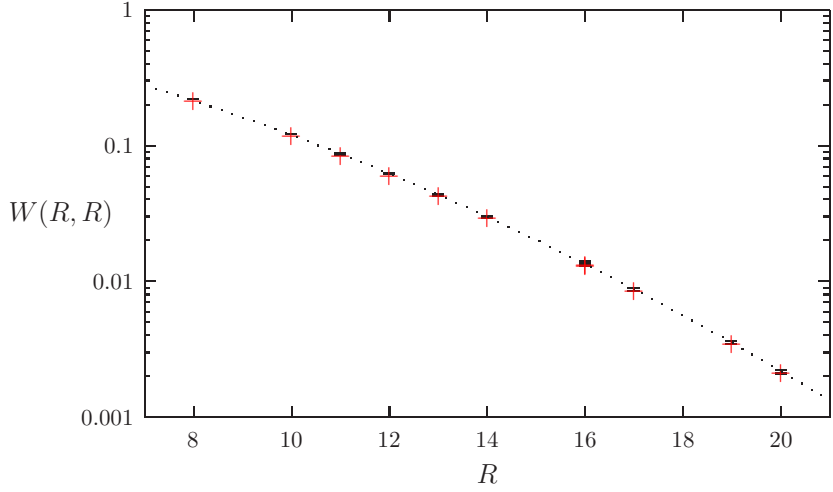}\\ 
%\vspace{5cm}  % amount of vertical space needed
\caption{Improved square Wilson loops evaluated using duality
  relation (\ref{dual}) and the projection on the largest FK cluster.
The simulation has been performed on a $40^3$ cubic lattice using 
$3.0\times 10^5$ Monte Carlo configurations.}
\label{fig:wilson_loop_fk}
\end{figure} 
\vskip .5 cm

 As we have anticipated, even in this model, like in the other
 gauge systems coupled to matter analysed up to now, the Wilson loop fails to
 exhibit string breaking in the normal, confining phase, due to its 
poor (or vanishing) projection  onto the screened potential. 
 One has to use additional operators with good projection on the
ground state, as first observed in Ref.\cite{Michael:1991nc}.
The static potential $V(R)$ and its excitations can be extracted from
measurements of the matrix correlator, represented pictorially as 

\vskip .2 cm
\begin{figure}[h]
%\begin{center}
%\input{largest.tex}
%\end{center}
\centering
\includegraphics[width=.5\textwidth]{cmatrix.epsi}\\ 
\end{figure}

where $C_{11}(R,T)$ is the rectangular Wilson loop $\bra W(R,T)\ket$,
the $U$-shaped operator is 
\eq
C_{12}(R,T)=\bra \varphi(0)U(0,-\vec{j}T)U(-\vec{j}T,
 -\vec{j}T+\vec{k}R)U(\vec{k}R-\vec{j}T,\vec{k}R)\varphi(\vec{k} R)
\ket~,
\label{op1}
\en
where $\vec{j}$ and $\vec{k}$ are two orthogonal unit vectors, and
$U(x,y)$ is a shorthand notation for a straight line of $U_\ell$
connecting the sites $x$ and $y$. The correlator is symmetric, 
$C_{12}(R,T)=C_{21}(R,T)$, and 
\eq
C_{22}(R,T)=\bra\varphi(0)U(0,\vec{j}T)\varphi(\vec{j}T)
\varphi(\vec{k}R)
U(\vec{k}R,\vec{j}T+\vec{k}R)\varphi(\vec{j}T+\vec{k}R)
\ket\;.
\label{op2}
\en
Denoting by $\lambda_0\leq\lambda_1$ the two eigenvalues, the ground
state potential is defined as
\eq V(R)=-\lim_{T\to\infty}\frac1T\log\lambda_0
\en 
In practical simulations, the limit $T\to\infty$ is not realized. In
our case $T$ was typically less that twenty lattice spacings when the 
the signal was lost in noise. 

The first point chosen in the present 
investigation is given by the couplings $\beta_G=0.75245$ and 
$\beta_I=0.16683$ . This point in the phase diagram of
Fig.\ref{fig:vacua} is in
the phase I, fairly close to the deconfinement transition, where the
model has properties very similar to the confining phase of pure gauge theory 
\cite{Gliozzi:2000yg}. The estimates of the above three operators in a
lattice of size $40^3$ as functions of $R$ are reported in 
Fig.\ref{fig:enl_basis}a and 
the static potential obtained the measure of 6 different $T$ in
Fig.\ref{fig:enl_basis}b.
We observe the typical
string breaking phenomenon with a Wilson loop which seems to follow
the area-law decay even in the region where the string is broken. In
order to see better this property, we used a powerful algorithm based
on Eq.(\ref{wc}) and already used in pure gauge theory
\cite{Caselle:1996ii}.
 We also modified each configuration in the Monte Carlo ensemble 
by eliminating  all the FK clusters not belonging to the largest
cluster. It has been shown that  this transformation does not change 
the value of the string tension but greatly reduce the noise of the 
measurements. The results are reported in Fig.\ref{fig:wilson_loop_fk}
: the square Wilson 
loops obey a perfect area-law even at  distances  five times the
string breaking scale at this point. 
\par
In order to investigate the effect of the electric condensate on the
Wilson loop, we considered two different points in the phase diagram,
one in the region I, at $\beta_G=0.6867$ and $\beta_I=0.20$, which is a 
normal confining phase. The other is
in the region II, at  $\beta_G=0.66$ and $\beta_I=0.27$ where besides
the CV (and the FK) infinite cluster in the
dual lattice there is also a percolating FK cluster in the direct
lattice, associated to the ``electric'' condensation. These two points are
chosen in such a way that the total size of the CV vortices are the
same (within the statistical errors) in the two cases. It turned out
that also the size of the maximal CV cluster is approximately the
same. We also verified that this quantity scales with the volume 
for large enough lattices. 

\par
\begin{figure}
%\begin{center}
%\input{largest.tex}
%\end{center}
\centering
\includegraphics[width=0.8\textwidth]{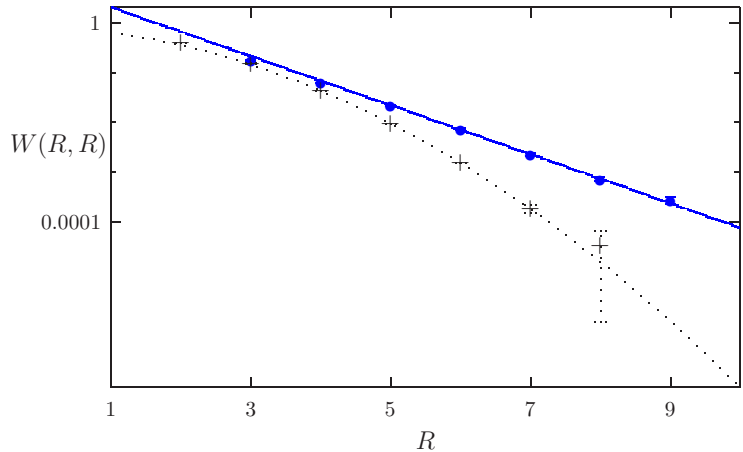} 
%\vspace{5cm}  % amount of vertical space needed
\caption{Square Wilson loops in the torn phase (circles) and in the
  normal, confining phase (crosses) resulting from two simulations in
  a cubic lattice of size $40^3$ with $2.0\times 10^5$ Monte Carlo
  configurations. The couplings are chosen in such
  a way that the density of center vortices is the same in the two
  cases.  The straight line is a fit to a perimeter-law decay, while
  the dotted line fits an area-law decay with $\sigma=0.1541(2)$.}
\label{fig:wilson_loop_phaseV}
\end{figure}

 
The dramatic effect produced by the electric condensate  is
demonstrated by the completely different behaviors of the square
Wilson loop in the two cases, as is evident in 
Fig.\ref{fig:wilson_loop_phaseV}. In the region I
we see again the characteristic area-law of the normal confining
phase, while the data of region II are compatible, for the Wilson
squares of larger size, with a decay with the perimeter. This is the
behavior expected for the torn phase \cite{gp}.

\section{Conclusions}
In this study we proposed to use the gauge theories coupled to
charged matter to test the confinement ideas of 't Hooft and
Mandelstam. In particular, in the case of the $Z_2$ gauge-Higgs 
model, we gave a detailed microscopic description of the center vortex
and monopole condensates. The analysis of the different vacua of this theory 
led us to conclude that
\begin{itemize}
\item 
confinement requires that both magnetic monopoles  
\underline{and} center vortices condense.  
\item
the condensation of the former does not imply necessarily  
the condensation of the latter: there are vacua with a percolating
FK cluster (i. e. the microscopic description of the magnetic condensate) 
which is not associated to an infinite CV cluster.
\item plausibility arguments suggest that these two kinds of infinite
  clusters are responsible of the area-law decay of Wilson loop only
  if the links of these clusters are weekly correlated.
\item there are vacua where the condensation of the gauge degrees 
of freedom (monopoles or center vortices) is associated with a
condensate of the matter field (the Higgs field in our example). This 
yields  strong correlations among vortices and monopoles so that the
plausibility arguments for confinement we alluded above are no longer
justified. In fact we observed a perimeter-law decay of large Wilson
loops. This could be identified with the torn phase described in 
\cite{gp}. 
\end{itemize}


\begin{thebibliography}{9}
%%%%%%%%%%%%%%%%%%%%%%%%%%%%%%%%%%%%%%%%%%%%%%%%%%%%%%%%%%%%%%%%%%%%%%%%%%
%
\bibitem{th}
G.~'t Hooft,
``On The Phase Transition Towards Permanent Quark Confinement,''
Nucl.\ Phys.\ B {\bf 138}, 1 (1978).
%%CITATION = NUPHA,B138,1;%%
%
\bibitem{tm}
G.~'t Hooft,
``Gauge Fields With Unified Weak, Electromagnetic, And Strong Interactions,''
in {\it C75-06-23.41}
Print-75-0836 (UTRECHT)
{\it Rapporteur's talk given at Int. Conf. on High Energy Physics, Palermo, Italy, Jun 23-28, 1975};
G.~Parisi,
``Quark Imprisonment And Vacuum Repulsion,''
Phys.\ Rev.\ D {\bf 11} (1975) 970;
S.~Mandelstam,
``Vortices And Quark Confinement In Nonabelian Gauge Theories,''
Phys.\ Rept.\  {\bf 23} (1976) 245;
G.~'t Hooft,
``Topology Of The Gauge Condition And New Confinement Phases In Nonabelian Gauge Theories,''
Nucl.\ Phys.\ B {\bf 190} (1981) 455.
%
%%%%%%%%%%%%%%%%%%%%%%%%%%%%%%%%%%%%%%%%%%%%%%%%%%%%%%%%%%%%%%%%%%%%%%%%%%
%
\bibitem{po} 
A.~M.~Polyakov,
``Compact Gauge Fields And The Infrared Catastrophe,''
Phys.\ Lett.\ B {\bf 59} (1975) 82, 
and 
``Quark Confinement And Topology Of Gauge Groups,''
Nucl.\ Phys.\ B {\bf 120} (1977) 429.
%
%%%%%%%%%%%%%%%%%%%%%%%%%%%%%%%%%%%%%%%%%%%%%%%%%%%%%%%%%%%%%%%%%%%%%%%%%%
%
\bibitem{Nakamura:1996sw}
N.~Nakamura, V.~Bornyakov, S.~Ejiri, S.~i.~Kitahara, Y.~Matsubara and 
T.~Suzuki,``Disorder parameter of confinement,''
Nucl.\ Phys.\ Proc.\ Suppl.\  {\bf 53} (1997) 512
[arXiv:hep-lat/9608004].
\bibitem{Chernodub:1996ps}
M.~N.~Chernodub, M.~I.~Polikarpov and A.~I.~Veselov,
``Effective constraint potential for Abelian monopole in SU(2) lattice
  gauge theory,''
Phys.\ Lett.\ B {\bf 399} (1997) 267
[arXiv:hep-lat/9610007].

\bibitem{ht}
A. Hart and M. Teper,
``Monopole clusters, Z(2) vortices, and confinement in SU(2),''
Phys.\ Rev.\ D {\bf 60} (1999) 114506
[arXiv:hep-lat/9902031].
%
%%%%%%%%%%%%%%%%%%%%%%%%%%%%%%%%%%%%%%%%%%%%%%%%%%%%%%%%%%%%%%%%%%%%%%%%%%
%
%\cite{DiGiacomo:1999fa}
\bibitem{DiGiacomo:1999fa}
A.~Di Giacomo, B.~Lucini, L.~Montesi and G.~Paffuti,
``Colour confinement and dual superconductivity of the vacuum. I,''
Phys.\ Rev.\ D {\bf 61} (2000) 034503
[arXiv:hep-lat/9906024].
%%CITATION = HEP-LAT 9906024;%%
%
%%%%%%%%%%%%%%%%%%%%%%%%%%%%%%%%%%%%%%%%%%%%%%%%%%%%%%%%%%%%%%%%%%%%%%%%%%
%
%\cite{DelDebbio:1996mh}
\bibitem{DelDebbio:1996mh}
L.~Del Debbio, M.~Faber, J.~Greensite and S.~Olejnik,
``Center dominance and Z(2) vortices in SU(2) lattice gauge theory,''
Phys.\ Rev.\ D {\bf 55} (1997) 2298
[arXiv:hep-lat/9610005].
%%CITATION = HEP-LAT 9610005;%%
%
%%%%%%%%%%%%%%%%%%%%%%%%%%%%%%%%%%%%%%%%%%%%%%%%%%%%%%%%%%%%%%%%%%%%%%%%%%
%
%\cite{Gliozzi:2001tu}
\bibitem{Gliozzi:2001tu}
F.~Gliozzi and A.~Rago,
``String breaking mechanisms induced by magnetic and electric  condensates,''
Nucl.\ Phys.\ Proc.\ Suppl.\  {\bf 106} (2002) 682
[arXiv:hep-lat/0110064].
%%CITATION = HEP-LAT 0110064;%%
%
%%%%%%%%%%%%%%%%%%%%%%%%%%%%%%%%%%%%%%%%%%%%%%%%%%%%%%%%%%%%%%%%%%%%%%%%%%
%
%\cite{Gliozzi:2002ht}
\bibitem{Gliozzi:2002ht}
F.~Gliozzi, M.~Panero and P.~Provero,
``Large center vortices and confinement in 3D Z(2) gauge theory,''
arXiv:hep-lat/0204030.
%%CITATION = HEP-LAT 0204030;%%
%
%%%%%%%%%%%%%%%%%%%%%%%%%%%%%%%%%%%%%%%%%%%%%%%%%%%%%%%%%%%%%%%%%%%%%%%%%%
%
%\cite{Engelhardt:1999fd}
\bibitem{Engelhardt:1999fd}
M.~Engelhardt, K.~Langfeld,H.~Reinhardt and O.~Tennert, 
``Deconfinement in SU(2) Yang-Mills theory as a center vortex
percolation  transition,''
Phys.\ Rev.\ D {\bf 61} (2000) 054504
[arXiv:hep-lat/9904004].
%%CITATION = HEP-LAT 9904004;%%
%
%%%%%%%%%%%%%%%%%%%%%%%%%%%%%%%%%%%%%%%%%%%%%%%%%%%%%%%%%%%%%%%%%%%%%%%%%%
%
\bibitem{Gliozzi:2002am}
F.~Gliozzi, M.~Panero and P.~Provero,
``Center vortices, magnetic condensate and confinement in a simple
gauge  system,''
arXiv:hep-lat/0205004.
%%CITATION = HEP-LAT 0205004;%%
%
%%%%%%%%%%%%%%%%%%%%%%%%%%%%%%%%%%%%%%%%%%%%%%%%%%%%%%%%%%%%%%%%%%%%%%%%%%
%
\bibitem{qcd}
B.~Bolder {\it et al.},
``A high precision study of the Q anti-Q potential from Wilson loops in  the regime of string breaking,''
Phys.\ Rev.\ D {\bf 63} (2001) 074504
[arXiv:hep-lat/0005018].
%%CITATION = HEP-LAT 0005018;%%
%
%%%%%%%%%%%%%%%%%%%%%%%%%%%%%%%%%%%%%%%%%%%%%%%%%%%%%%%%%%%%%%%%%%%%%%%%%%
%
%\cite{Michael:1991nc}
\bibitem{Michael:1991nc}
C.~Michael,
``Hadronic forces from the lattice,''
Nucl.\ Phys.\ Proc.\ Suppl.\  {\bf 26} (1992) 417.
%%CITATION = NUPHZ,26,417;%%
[arXiv:hep-ph/9809211].
%
%%%%%%%%%%%%%%%%%%%%%%%%%%%%%%%%%%%%%%%%%%%%%%%%%%%%%%%%%%%%%%%%%%%%%%%%%%
%
\bibitem{ks}O.~Philipsen and H.~Wittig,
``String breaking in non-Abelian gauge theories with fundamental matter  fields,''
Phys.\ Rev.\ Lett.\  {\bf 81} (1998) 4056
[Erratum-ibid.\  {\bf 83} (1999) 2684]
[arXiv:hep-lat/9807020];
%%CITATION = HEP-LAT 9807020;%%
F.~Knechtli and R.~Sommer  [ALPHA collaboration],
``String breaking in SU(2) gauge theory with scalar matter fields,''
Phys.\ Lett.\ B {\bf 440} (1998) 345
[arXiv:hep-lat/9807022];
%%CITATION = HEP-LAT 9807022;%%
F.~Knechtli and R.~Sommer  [ALPHA Collaboration],
``String breaking as a mixing phenomenon in the SU(2) Higgs model,''
Nucl.\ Phys.\ B {\bf 590} (2000) 309
[arXiv:hep-lat/0005021].
%%CITATION = HEP-LAT 0005021;%%
%
%%%%%%%%%%%%%%%%%%%%%%%%%%%%%%%%%%%%%%%%%%%%%%%%%%%%%%%%%%%%%%%%%%%%%%%%%%
%
\bibitem{milc}
C.~W.~Bernard {\it et al.},
``Zero temperature string breaking in lattice quantum chromodynamics,''
Phys.\ Rev.\ D {\bf 64} (2001) 074509
[arXiv:hep-lat/0103012].
%%CITATION = HEP-LAT 0103012;%%
%
%%%%%%%%%%%%%%%%%%%%%%%%%%%%%%%%%%%%%%%%%%%%%%%%%%%%%%%%%%%%%%%%%%%%%%%%%%
%
\bibitem{gp}
F.~Gliozzi and P.~Provero,
``The confining string and its breaking in QCD,''
Nucl.\ Phys.\ B {\bf 556} (1999) 76
[arXiv:hep-lat/9903013] and
%%CITATION = HEP-LAT 9903013;%%
%F.~Gliozzi and P.~Provero,
%``When QCD strings can break,''
Nucl.\ Phys.\ Proc.\ Suppl.\  {\bf 83} (2000) 461
[arXiv:hep-lat/9907023].
%%CITATION = HEP-LAT 9907023;%%
%
%%%%%%%%%%%%%%%%%%%%%%%%%%%%%%%%%%%%%%%%%%%%%%%%%%%%%%%%%%%%%%%%%%%%%%%%%%
%
\bibitem{Bertle:2001ya}
R.~Bertle, M.~Faber and A.~Hirtl,
``Vortices in the SU(2)-Higgs model: Vortices and the covariant
adjoint  Laplacian,''
Nucl.\ Phys.\ Proc.\ Suppl.\  {\bf 106} (2002) 664
[arXiv:hep-lat/0110098].
%%CITATION = HEP-LAT 0110098;%%
%
%%%%%%%%%%%%%%%%%%%%%%%%%%%%%%%%%%%%%%%%%%%%%%%%%%%%%%%%%%%%%%%%%%%%%%%%%%
%
\bibitem{fk}
C.~M.~Fortuin and P.~W.~Kasteleyn,
``On The Random Cluster Model. 1. Introduction And Relation To Other Models,''
Physica {\bf 57} (1972) 536.
%%CITATION = PHYSA,57,536;%%
%
%%%%%%%%%%%%%%%%%%%%%%%%%%%%%%%%%%%%%%%%%%%%%%%%%%%%%%%%%%%%%%%%%%%%%%%%%%
%
\bibitem{gv}
F.~Gliozzi and S.~Vinti,
``Nature of the vacuum inside the color flux tube,''
Nucl.\ Phys.\ Proc.\ Suppl.\  {\bf 53} (1997) 593
[arXiv:hep-lat/9609026].
%%CITATION = HEP-LAT 9609026;%%
%
%%%%%%%%%%%%%%%%%%%%%%%%%%%%%%%%%%%%%%%%%%%%%%%%%%%%%%%%%%%%%%%%%%%%%%%%%%
%
%\cite{Caselle:1999hy}
\bibitem{Caselle:1999hy}
M.~Caselle and F.~Gliozzi,
``Thermal Operators in Ising Percolation,''
J.\ Phys.\ A {\bf 33} (2000) 2333
[arXiv:cond-mat/9905234].
%%CITATION = COND-MAT 9905234;%%
%
%%%%%%%%%%%%%%%%%%%%%%%%%%%%%%%%%%%%%%%%%%%%%%%%%%%%%%%%%%%%%%%%%%%%%%%%%%
%
\bibitem{js}
G.~A.~Jongeward and J.~D.~Stack,
``Monte Carlo Calculations on Z(2) Gauge - Higgs Theories,''
Phys.\ Rev.\ D {\bf 21} (1980) 3360.
%%CITATION = PHRVA,D21,3360;%%
%
%%%%%%%%%%%%%%%%%%%%%%%%%%%%%%%%%%%%%%%%%%%%%%%%%%%%%%%%%%%%%%%%%%%%%%%%%%
%
%\cite{Gliozzi:2000yg}
\bibitem{Gliozzi:2000yg}
F.~Gliozzi,
``Some universal properties of the string breaking,''
Nucl.\ Phys.\ Proc.\ Suppl.\  {\bf 94} (2001) 550
[arXiv:hep-lat/0010084].
%%CITATION = HEP-LAT 0010084;%%
%
%%%%%%%%%%%%%%%%%%%%%%%%%%%%%%%%%%%%%%%%%%%%%%%%%%%%%%%%%%%%%%%%%%%%%%%%%%
%
%\cite{Caselle:1996ii}
\bibitem{Caselle:1996ii}
M.~Caselle, R.~Fiore, F.~Gliozzi, M.~Hasenbusch and P.~Provero,
``String effects in the Wilson loop: A high precision numerical test,''
Nucl.\ Phys.\ B {\bf 486} (1997) 245
[arXiv:hep-lat/9609041].
%%CITATION = HEP-LAT 9609041;%%
%
\end{thebibliography}
\end{document}